\documentclass[conference]{IEEEtran}
\IEEEoverridecommandlockouts
\usepackage{cite}
\usepackage{float}
\usepackage{amsmath,amssymb,amsfonts,mathtools}
\usepackage{algpseudocode}
\usepackage{algorithm}
\usepackage{array,multirow,graphicx}
\usepackage{textcomp}
\usepackage{subfig}
\usepackage{booktabs}
\usepackage{enumitem}
\usepackage{threeparttable}
\usepackage{todonotes}
\usepackage[multiple]{footmisc}
\usepackage{balance}
\usepackage{xcolor,colortbl,soul}
\usepackage[hyphens]{url}

\def\BibTeX{{\rm B\kern-.05em{\sc i\kern-.025em b}\kern-.08em
    T\kern-.1667em\lower.7ex\hbox{E}\kern-.125emX}}

\usepackage[hyperfootnotes=false]{hyperref}

\definecolor{Gray}{gray}{0.85}
\newcolumntype{a}{>{\columncolor{Gray}}c}

\sethlcolor{lightgray}

\begin{document}

\title{Phoebe: QoS-Aware Distributed Stream Processing through Anticipating Dynamic Workloads\\
}

\author{
\IEEEauthorblockN{
Morgan K. Geldenhuys\IEEEauthorrefmark{1}, 
Dominik Scheinert\IEEEauthorrefmark{1}, 
Odej Kao\IEEEauthorrefmark{1}, and 
Lauritz Thamsen\IEEEauthorrefmark{3}
}
\IEEEauthorblockA{
\IEEEauthorrefmark{1}
Technische Universit{\"a}t Berlin, Germany, \{firstname.lastname\}@tu-berlin.de}
\IEEEauthorblockA{
\IEEEauthorrefmark{3}
University of Glasgow, United Kingdom, lauritz.thamsen@glasgow.ac.uk}
}

\maketitle

\begin{abstract}

Distributed Stream Processing systems have become an essential part of big data processing platforms. 
They are characterized by the high-throughput processing of near to real-time event streams with the goal of delivering low-latency results and thus enabling time-sensitive decision making.
At the same time, results are expected to be consistent even in the presence of partial failures where exactly-once processing guarantees are required for correctness.
Stream processing workloads are oftentimes dynamic in nature which makes static configurations highly inefficient as time goes by.
Static resource allocations will almost certainly either negatively impact upon the Quality of Service and/or result in higher operational costs.

In this paper we present Phoebe, a proactive approach to system auto-tuning for Distributed Stream Processing jobs executing on dynamic workloads. 
Our approach makes use of parallel profiling runs, QoS modeling, and runtime optimization to provide a general solution whereby configuration parameters are automatically tuned to ensure a stable service as well as alignment with recovery time Quality of Service targets.
Phoebe makes use of Time Series Forecasting to gain an insight into future workload requirements thereby delivering scaling decisions which are accurate, long-lived, and reliable.
Our experiments demonstrate that Phoebe is able to deliver a stable service while at the same time reducing resource over-provisioning.

\end{abstract}


\section{Introduction}
\label{sec:introduction}

Distributed Stream Processing (DSP) systems are responsible for extracting valuable insights from large streams of real-time data.
Application areas include IoT data processing, click stream analysis, network monitoring, fraud detection, spam filtering, news processing, and many more\cite{Nasiri2018ASO, IAM+19, NNG19}.
It is here where high throughput rates, low end-to-end latencies, and the ability to continue operating in the presence of partial failures are essential for supporting time-sensitive decision making.
Input streams, however, are dynamic in nature and processing workloads, therefore, have the potential to change significantly over time.
Consequently, DSP systems like Apache Flink\cite{CKE+15}, Spark\cite{ZCF+10}, and Heron\cite{Kulkarni2015TwitterHS} are able to scale horizontally across a cluster of commodity nodes in order to accommodate variable processing loads.
At the same time, the complexity with which these systems are composed makes the manual optimization of configuration parameters a challenging task, even for experts.
The heterogeneous nature of these environments makes using a one-size-fits-all approach to configuration highly inefficient whereas finding individually optimized setups is difficult and time consuming.
This inevitably results in users prioritizing the over-provisioning of resources over lower operational costs to ensure DSP jobs operate within the bounds of expected Quality of Service (QoS) constraints.

Consequently, we have seen the introduction of research focused on providing horizontal autoscaling for DSP environments\footnote{\url{https://flink.apache.org/2021/05/06/reactive-mode}, Accessed: Mar 2022}\footnote{\url{https://databricks.com/session/auto-scaling-systems-with-elastic-spark-streaming}, Accessed: Mar 2022}~\cite{GSH+14, FAG+17, PETROV2018109, HuKZ19, KLH+18}.
This generally involves the automatic tuning of configuration parameters at runtime in an attempt to improve overall operational efficiency by matching computational resources to the current workload requirements.
The majority of existing approaches rely on course-grained metrics to make scaling decisions usually involving monitoring for bottlenecks and fixed/percentage-based resource adjustments. 
Further approaches model the scaleout behaviors of DSP jobs in an attempt to improve the accuracy of scaling decisions.
Irrespective of performance, most ignore \emph{exactly-once}\footnote{\url{https://www.confluent.io/blog/exactly-once-semantics-are-possible-heres-how-apache-kafka-does-it/}, Accessed: Mar 2022} processing requirements and none to the best of our knowledge have as yet taken recovery time planning into consideration.
For Checkpoint and Rollback Recovery (CPR), downtime is directly related to resource allocations where the amount of resources in excess of what is used to keep up with normal workload rates determines how quickly a job can recover after failure.
As a result, targeting 100\% utilization for normal workloads would cause the DSP job to permanently be stuck in the recovery phase while falling further and further behind processing events being produced at the latest timestamp.

In order to provide a stable service which produces near-optimal performance, solutions should encapsulate the following characteristics not fully embodied by existing approaches:

\begin{itemize}[leftmargin=*]
    \setlength{\itemsep}{3pt}
    
    \item{\verb|Accurate scaling decisions|}: The autoscaler should be able to select scaleouts which result in near-optimal end-to-end processing latencies at any specific workload while ensuring high utilization of allocated resources. Also, to provide a generalizable solution, DSP jobs should be regarded as operating as a black box.
    
    \item{\verb|Long-lived reconfigurations|}: Re-scaling a running job is expensive. Stateful processing pipelines where task instances are not completely independent are unable to dynamically re-scale the number of tasks during execution while still guaranteeing exactly-once processing. Reconfigurations require controlled restarts and therefore introduce interruptions to the service. 
    As such, it is imperative to ensure the number of restarts is kept to an absolute minimum.

    \item{\verb|Reliable recovery times|}: In CPR, jobs are essentially unavailable from the time a failure occurs to the point when the job has once again caught-up to processing events at the latest timestamp. Therefore, in the presence of QoS recovery time targets, it is important to allocate adequate resources ensuring recovery is within acceptable limits.
\end{itemize}

In this paper we present Phoebe, a proactive approach to system auto-tuning for DSP jobs executing in cloud environments where streaming workloads are expected to change over time.
It achieves this by executing the following procedure: firstly, orchestrating parallel profiling runs whereby the effects of varying resource allocations across changing throughput rates are measured; next, results are used to train two models for predicting average end-to-end latencies and recovery times; and lastly, an iterative optimization step where the scaleout of targeted DSP jobs is optimized based on these models in relation to dynamic workloads.
Profiling runs are conducted once, after which the resulting models are continuously updated with runtime metrics to ensure alignment with current cluster conditions.
Phoebe makes use of Time Series Forecasting (TSF) to gain an insight into future workload requirements and delivers a proactive approach to autoscaling which embodies the aforementioned characteristics.
We implemented Phoebe prototypically together with Apache Flink we demonstrate its usefulness in comparison with two state of the art approaches.

\section{Background}

This section expands upon the background related to DSP systems, CPR, failure types, and horizontal autoscalers.

\subsection{Distributed Stream Processing Systems}

DSP systems are inherently parallel and distributed. Streaming jobs are composed as a dataflow graph where task vertices contain some user-defined logic and streaming edges pass messages between the tasks \cite{SCS17}. 
They are, in principle, required to operate indefinitely on an unbounded stream of continuous data to produce new results as elements enter the streams. 
Because streams are unbounded, aggregating events works differently than in batch processing. Here, in order to perform some statistical analysis, aggregates are scoped by \textit{windows}. 
Windows can either be data-driven, i.e. requiring a certain number of events to progress, or time-driven. 
Time is an important notion in stream processing which has ramifications for both performance and the consistency of results. 
Although three separate times can be distinguished in streaming jobs, we only consider \textit{event time} processing. 
This is where a timestamp is extracted from each event and together with watermarks are used for time-based operations, i.e. the progress of time depends on the data, not on any wall clocks. It has the most impact on performance when compared to the other times, but events are allowed to arrive out-of-order and results are consistent even in the presence of failures\footnote{Flink: Event Time. URL: \url{https://ci.apache.org/projects/flink/flink-docs-stable/dev/event\_time.html}, Accessed: Mar 2022}.

\subsection{Checkpoint and Rollback Recovery}

The \textit{global state} of a DSP system is represented by the instantaneous local state at each process as well as the instantaneous state of the channels, i.e. the messages in transit. 
Recording the global state is an important paradigm which in the context of fault tolerance allows for the system to be recovered when failures occur. 
This is achieved by periodically saving, called \emph{checkpointing}, and then restoring the system to the last saved global state. Recording on-the-fly global states of a distributed system where there are no bounds on message delays, referred to as the snapshot problem, is non-trivial. 
This is mainly due to both the lack of globally shared memory and a global clock making synchronization between nodes extremely difficult \cite{KRS95} \cite{FLP85}. 
Thankfully synchronization is not required, preserving causality within a distributed system is enough for recording a consistent global snapshot \cite{FLP85}. 
This forms the basis of the \textit{Chandy-Lamport algorithm} upon which checkpoint and rollback recovery strategies are based \cite{CL85}. 
Its goal is to record a set of process and channel states such that, even though the combination of states may never have occurred at the same time, the recorded global state is consistent. 
Another aspect which needs to be considered is the consistency of results across a pipeline of interacting systems and the effect that events will have on the global state due to failures. 
For this, certain systems provide fault tolerance guarantees. 
In this paper we consider the strongest of these, \textit{exactly-once} delivery, where each event is guaranteed to pass through the system once and only once and results are ensured to be correct even in the presence of failures. 
This, however, introduces overhead that impacts on performance.

\begin{figure*}[!h]
    \centering
    \includegraphics[width=1.0\textwidth]{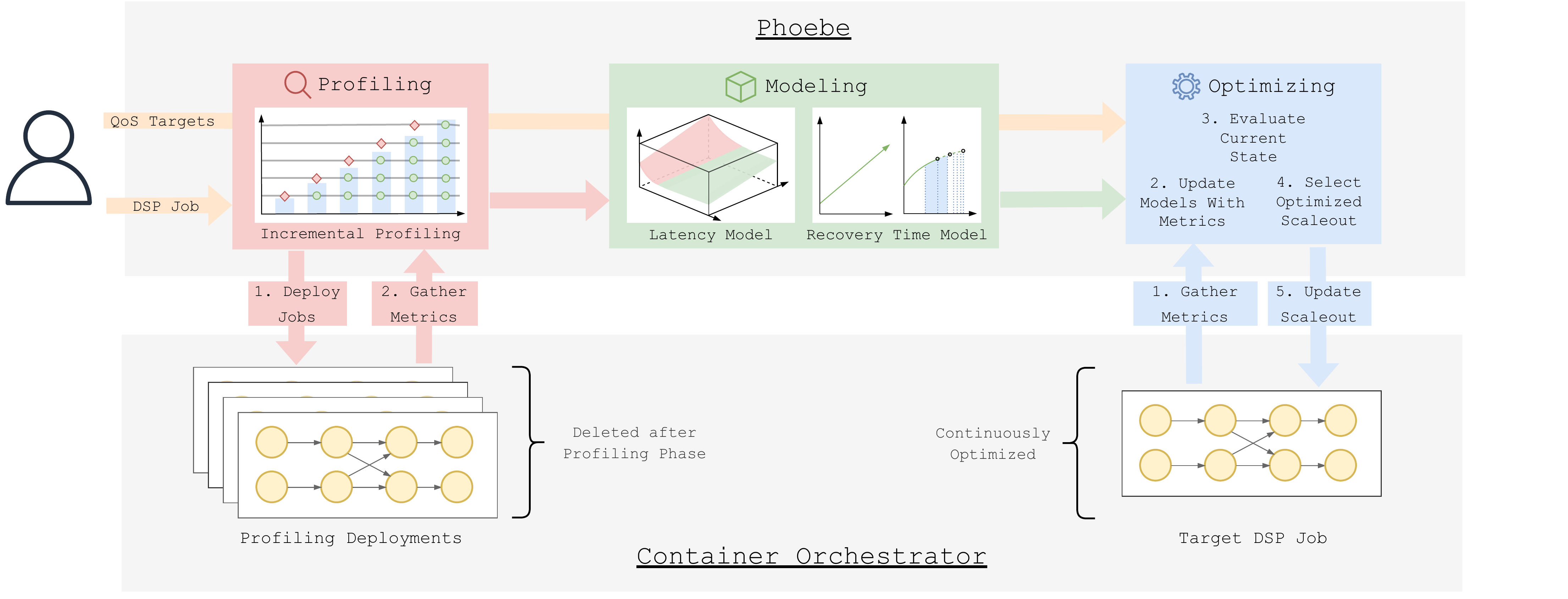}
    \caption{High level overview of Phoebe depicting three phases of the approach including interactions with users and systems.}
    \label{fig:overview}
\end{figure*}

\subsection{Horizontal Autoscaling}

Most modern DSP systems~\cite{CKE+15,ZCF+10} offer the functionality to reconfigure the system in such a way that the number of tasks assigned to each running job can be changed after processing has already begun.
Such functionality is necessary as optimizing the scaleout can reduce both cost and resource wastage which comes with its own sustainability concerns.
In cases where the consistency of results must be guaranteed, reconfigurations are usually accomplished by using the CPR mechanism where the global state of the running job is saved, the number of tasks are adjusted based on new workload requirements, and the job is restarted from the latest checkpoint.
The ability to manage these reconfigurations automatically, however, is not a core capability of DSP systems and is generally left up to the user to perform manually based on their expert knowledge.
For this reason, horizontal autoscalers have been proposed which can automatically adjust the number of tasks available to a running DSP job. 
It does so by observing some physical runtime property/properties and comparing it/them to a predefined desired state.
Examples of such properties include CPU and/or memory utilization, end-to-end latencies, throughput rates, backpressure, recovery times, etc.
The autoscaler is thus able to infer the near-optimal parallelism which will result in conditions returning close to the desired state and subsequently initiate a reconfiguration.

\section{Approach}
\label{sec:approach}

The goal of Phoebe is to optimize performance as well as availability for targeted DSP jobs operating on dynamic workloads.
Phoebe does this by providing a more efficiently executing DSP system which maximizes utilization by reducing over- and under-provisioning of resources.
Our approach therefore optimizes the \emph{scaleout} ($s$) configuration variable in relation to the changing workload requirements.
Performance and availability are defined in terms of \emph{average end-to-end latency} ($L_{avg}$) and \emph{recovery time} ($R$) respectively.
The $L_{avg}$ is defined as the average time required for events to traverse the execution graph of the DSP job from the point when measurements are taken until they arrive at the sink operators. 
This is the response time of the system considering event-time processing and excludes windowing periods.
$R$ refers to the time required for the job to catch up to processing events being produced close to the latest timestamp after a failure and rollback recovery.
The user is required to define a parameter $RC_{tar}$ which defines the recovery time to optimize for after which the DSP job should once again be available.

\subsection{General Idea}

To achieve its goal, three distinct phases are executed sequentially for each targeted job.
The first is a profiling phase described in~\autoref{sec:profiling} where live data is recorded, then short-lived parallel profiling jobs are generated and the data is replayed at various workload rates across various scaleouts.
The data produced from this initial phase is fed into the second phase described in~\autoref{sec:modeling} where two models are generated: an availability model used in predicting the scaleout needed to produce a recovery time close to $RC_{tar}$; and a performance model used for predicting the scaleout which will result in stable processing latencies based on the average workload rate.
Lastly, Phoebe uses these models to perform an online optimization step in the final phase described in~\autoref{sec:optimization} which evaluates the current state and future potential for performance and recovery time improvements.
In doing so, not only does it react to the current conditions but likewise anticipates future workload requirements and therefore proactively makes scaling decisions intended to reduce disruptions to the service.
The optimization phase is intended to execute indefinitely with the models being continuously updated to ensure continued accuracy as cluster conditions change over time.
An overview of the entire process can be seen in Fig. \ref{fig:overview}.
Before detailing the phases any further, a method for estimating recovery time is required to enable our approach.

\subsection{Estimating Recovery Time}
\label{sec:recovery_time_derivation}

When enforcing exactly-once processing guarantees, DSP jobs are essentially unavailable from the point when the failure occurs until it has once again caught-up to processing events at the latest timestamp.
CPR consists of two distinct phases:

\begin{itemize}[leftmargin=*]
    \setlength{\itemsep}{3pt}
    
    \item{\verb|Downtime| ($D$)}: Represents the length of time required to detect the failure and restart the job from the last successful checkpoint. 
    Generally, a heartbeat mechanism is used to determine when a remote process has stopped responding and the timeout is set via a configuration parameter.  
    After this phase, the job is once again able to process events.

    \item{\verb|Catch-up| ($C$)}: Represents the length of time required to catch-up to the latest timestamp from the point where the job is once again able to resume processing. 
    This involves processing the backlog of events accumulated while the job was unavailable from the last checkpoint as well as the events arriving while in the catch-up phase. 
    The DSP system will use the maximum processing capacity available in an attempt to catch-up and the more physical resources that are available, the faster this process will take to complete.
    
\end{itemize}

As restart times do not vary greatly across failures, an average can be assumed based on observations.
Consequently, a reasonably good estimate of $D$ can be made prior to job execution.
This is not the case for $C$, however, whose duration is directly related to when the last checkpoint was made, the time spent in $D$, and the future workload requirements.
Consequently, in order to estimate the \emph{recovery time} ($R$), a heuristic for calculating $C$ is needed based on how the throughput rate is most likely to change over time.
We formulate this as a geometrically decreasing sequence where the $1^{st}$ term predicts the time required to process the backlog of events accumulated while the job was forced to restart and each subsequent term then predicts the time required to process events arriving as a result of executing the previous step.
The input $n$ is an integer which represents the number of steps to be executed.
Therefore, using the \emph{maximum processing capacity} ($T_{max}$) of the system and assuming enough resources have been allocated to make progress, the duration of each step decreases as it approach the point at which events are being generated.
$T_{max}$ is based on the current scaleout.
We define this sequence as:

\begin{equation*}
    {
        c(n)= 
        \begin{cases}
            \displaystyle\int\limits_{a}^{b} f(t) \, \mathrm{d} t \cdot \frac{1}{T_{max}}, & 
                \begin{array}{@{}l@{}@{}}
                    \text{if $n=0$}\\
                    \text{where $a=t_{0}-I$}\\
                    \text{and \, \, $b=t_{0}+D$} 
                \end{array} \\ \addlinespace [0.5em]
            \displaystyle\int\limits_{a}^{b} f(t) \, \mathrm{d} t \cdot \frac{1}{T_{max}}, & 
                \begin{array}{@{}l@{}}
                    \text{if $n>0$}\\
                    \text{where $a=t_{n}$}\\
                    \text{and \, \, $b=t_{n}+c(n-1)$} 
                \end{array}
        \end{cases}
    }
\end{equation*}

\begin{figure}
    \centering
    \includegraphics[width=\columnwidth]{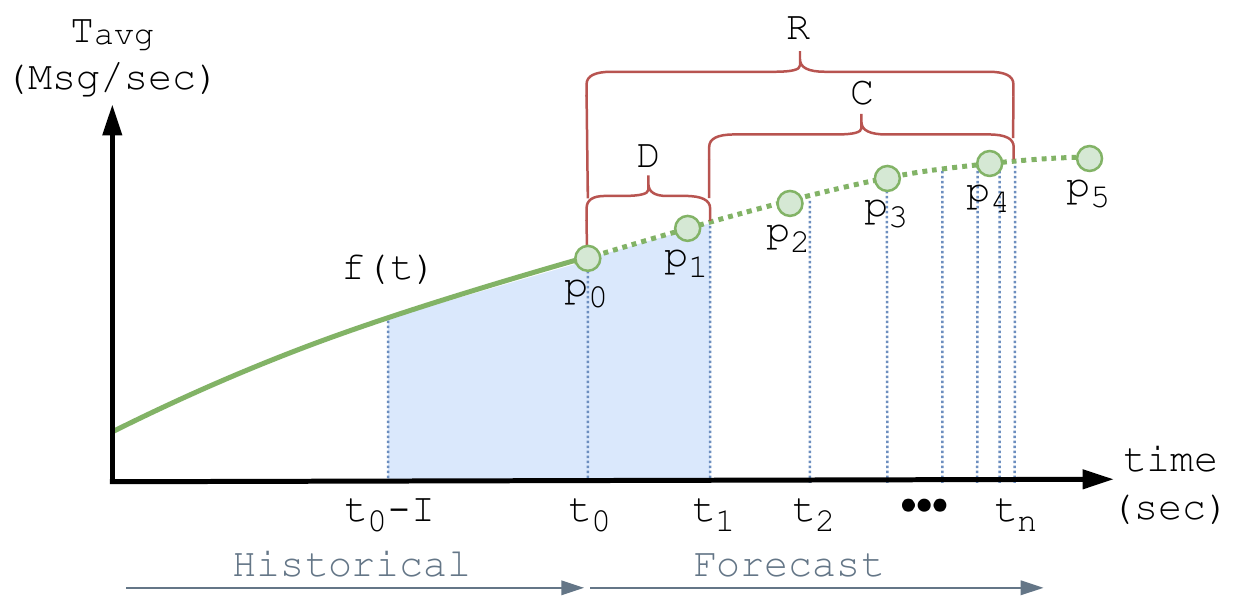}
    \caption{Depiction of runtime recovery time estimation for streaming jobs processing dynamic workloads.}
    \label{fig:recovery_time}
\end{figure}

We use multistep-ahead TSF\cite{Taieb2012ARA} to model how the workload is most likely to change over time.
This takes historical throughput rates as input and predicts a set of time steps up until a specific time horizon.
Fig. \ref{fig:recovery_time} provides a graphical representation of the recovery time estimation procedure.
$f$ is a time series and represents the function of the \emph{average throughput rate} (T\textsubscript{avg}) over time.
At the current timestamp $t_{0}$, the historical data is inputted into the chosen TSF algorithm to forecast the future workload.
Points $P_{0}$ to $P_{5}$ represent the outputted set of time steps which are then interpolated and combined with historical data to produce $f$.
The highlighted area under the curve between $t_0-I$ and $t_1$ represents the backlog of events for which the $1^{st}$ term of $c$ calculates the processing time based on the $T_{max}$ of the system.
$I$ in this context represents the timestamp of when the last checkpoint completed successfully and therefore the point at which processing will restart from.
This output of $1^{st}$ term is therefore equal to the time period between $t_{1}$ and $t_{2}$, i.e. the time it took to process the preceding step.
This process is repeated with the assumption that the subsequent processing times approach zero.
In order to calculate the recovery time the number of steps $n$ needs to be determined.
This is done iteratively by inputting increasing integer values into $c$ in order to find an appropriate $n$, i.e. the first $n$ where the last step of the sequence produces a value sufficiently close to zero, e.g. less than 1 second.
With the number of steps $n$ being determined, $C$ can be calculated by measuring the length of time between $t_{1}$ and $t_{n}$.
Finally, the time periods $D$ and $C$ are added together to give an estimation of $R$.

It should be noted that, assuming a constant $T_{max}$, the variability of $R$ is dependent on $I$.
For our approach we use the \emph{checkpoint interval} configuration variable for the value of $I$, i.e. we assume that failures occur just before the next checkpoint completes successfully. 
With this mechanism we can now proceed to describing the phases of our approach.

\subsection{Phase 1: Profiling}
\label{sec:profiling}

This initial phase is performed in order to extract metrics which are then used for modeling in the subsequent phase.
This is done through profiling parallel deployments of the targeted DSP job at differing scaleout configurations over increasing workload rates.
In this phase we are attempting to quantify how the $L_{avg}$ is impacted by scaling out as the workload changes and what is the $T_{max}$ of each deployment based on the scaleout.
Given user-defined minimum scaleout $S_{\min}$, maximum scaleout $S_{\max}$, and desired number of parallel profiling runs $S_{count}$, we construct a set $\mathbf{S}$ of equally spaced scaleouts in the given range:

\begin{equation*}
\mathbf{S} = \{S_{\min}, \ldots, S_{\max}\},\text{ s.t. } |\mathbf{S}| = S_{count}.
\end{equation*}

\begin{figure}
    \centering
    \includegraphics[width=\columnwidth]{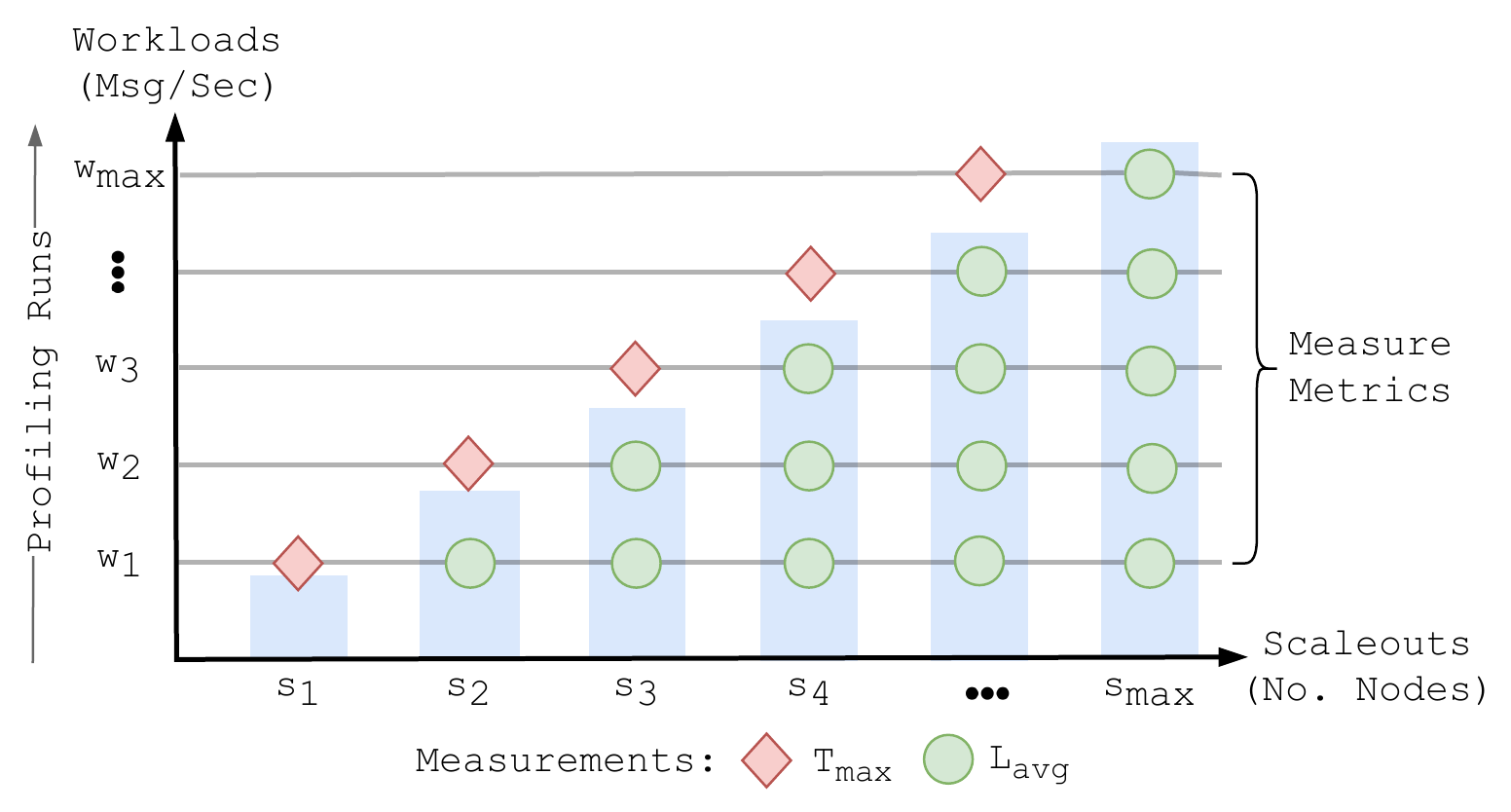}
    \caption{Measuring scaleout performance and maximum processing capacity over increasing workload rates.}
    \label{fig:profiling}
 \end{figure}

Next, it is required to generate a stream of events for profiling at increasing throughput rates.
To achieve this, Phoebe will connect to the streaming sources where the targeted DSP job is consuming input data and record events for the finite length of time.
We make the assumption that live streams of production-like data are available and the volume of incoming events is sufficient for profiling.
On startup, Phoebe connects to the cluster and creates for each scaleout $S_i \in \mathbf{S}$ an instance of the targeted DSP job with otherwise identical configuration.
Each deployment is configured to consume from the same source(s).
Next, a number of staggered profiling runs are conducted where Phoebe replays the recorded messages at increasing workload rates.
Phoebe is able to scale the rate of recorded events by randomly deleting or replicating existing events and therefore able to produce a constant stream of profiling data.
\autoref{fig:profiling} shows a graphical overview of the profiling scheme where shaded areas represent the $T_{max}$ corresponding to a deployment with scaleout $S_i\in \mathbf{S}$.
After each profiling run, $L_{avg}$ values are measured and evaluated to determine whether or not they are able to perform adequately.

Evaluating scaleout behaviors as it relates to $L_{avg}$ can be done using a combination of clustering and regression techniques.
If enough resources have been allocated, valid scaleouts will perform similarly well at the same $T_{avg}$.
Conversely, when resources are in shortfall, the $L_{avg}$ will increase over time as the job falls behind processing requirements. 
As such, this configuration should be considered invalid.
As long as there are more than two scaleouts to evaluate, clustering the results should produce at minimum one group of closely clustered values.
If two groups are produced, then the group with the lowest aggregate $L_{avg}$ should be considered valid with the other being invalid.
For fewer than three scaleouts, too few parallel data points exist to form a majority and so a regression line is modeled using the previously measured $L_{avg}$ values of each individual scaleout.
If the predicted value deviates greatly from the actual value, then the configuration is deemed invalid for the current profiling workload.

When an invalid $L_{avg}$ is detected, the maximum processing capacity at this specific scaleout has been found and $T_{max}$ is revealed.
The deployment in question has therefore outlived its usefulness and is deleted, thus releasing the cluster resources and reducing the cost of profiling.
In both cases $S_{i}$ and $T_{avg}$ are saved alongside these measurements.
Profiling will continue at incrementally increasing workload rates until $T_{max}$ of all parallel deployments is determined. 
Results are then passed into the next phase, namely the modeling phase.

\subsection{Phase 2: Modeling}
\label{sec:modeling}

As part of our approach we are concerned with achieving near-optimal performance in terms of $L_{avg}$ and planned recovery in terms of $R_{tar}$ for targeted DSP jobs.
To this end, we develop two models, $M_{L}$ and $M_{R}$, each utilizing and composing multiple techniques in order to achieve our goals.

As mentioned in~\autoref{sec:profiling}, during profiling we observe a distinct range of scaleouts over varying workload rates and measure the $L_{avg}$. 
Consequently, we train a multiple regression model $l_1$ for predicting $L_{avg}$ values based on scaleouts and workload rates as inputs.
However, predicting latencies in such a way does not fully solve the problem of finding a suitable scaleout as we do not yet know which set of latencies should be considered valid.
Our goal is to target the smallest scaleout producing a stable $L_{avg}$, i.e. unstable latencies are continuously increasing as not enough resources have been allocated to allow the DSP job to keep up with the workload requirements.
Therefore, in a method similar to what was used during profiling, we employ a clustering algorithm to identify two clusters of the predicted latencies.
The cluster with the smallest centroid is considered to contain the set of valid latencies.
We find this approach to be particularly effective when values $L_{avg}$ are preprocessed by normalizing with respect to the boundaries of the 1st percentile of the data, and thereafter applying a logarithmic transformation. 

\begin{figure}
    \centering
    \includegraphics[width=\columnwidth]{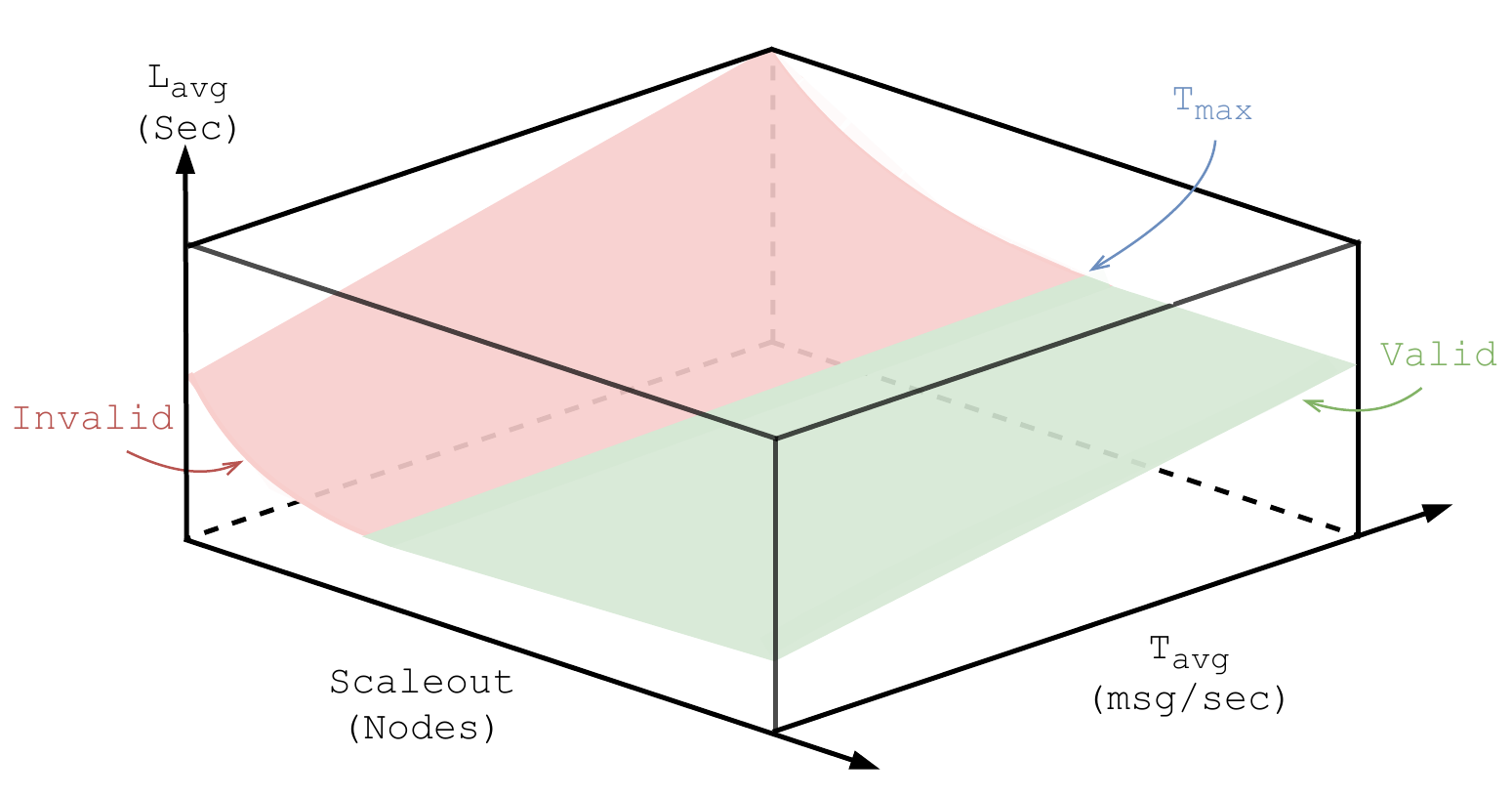}
    \caption{Model for predicting the average end-to-end latency based on scaleout and average input throughput rate.}
    \label{fig:performance}
\end{figure}

Therefore, with our model $M_{L}$ internally utilizing methods for both multiple regression and clustering, it allows for the estimation of latencies for any combination of scaleouts and workload rates, and delivers an indication about the goodness of predictions.
\autoref{fig:performance} provides a simplified graphical representation of how the resulting plane shows an increase in $S_i$ and/or a decrease in $T_{avg}$ will likewise decrease the $L_{avg}$.
Conversely, a decrease in $S_i$ and/or an increase in $T_{avg}$ will result in the $L_{avg}$ increasing.
It is important to note that the variability with which the $L_{avg}$ prediction is impacted is dependent on the DSP job and current cluster conditions.

For planned recovery, we follow a different approach.
As shown in~\autoref{sec:recovery_time_derivation}, estimating recovery times requires both knowledge of the $T_{max}$ of the DSP job as well as how the $T_{avg}$ is mostly likely to change over time. 
Therefore, a regression model is used to approximate the $T_{max}$ for any scaleout $S_i \in \mathbf{S}$ using the profiling data.
Then we utilize a multistep-ahead TSF model which is trained on historical data, specifically the workload rate over time.
We define a maximum TSF horizon of $H_{max}$.
Both methods are thus used together with current metrics to realize the estimation of recovery times.
Following our approach to recovery time estimation, $M_R$ can, in conjunction with the regression technique for $T_{max}$ as well as multistep-ahead TSF of the expected workload rate over time, estimate the recovery time $R$ for any scaleout $S_i \in \mathbf{S}$.
After the models are trained, optimization can begin.

\subsection{Phase 3: Optimization}
\label{sec:optimization}

\begin{figure}[]
    \centering
    \includegraphics[width=\columnwidth]{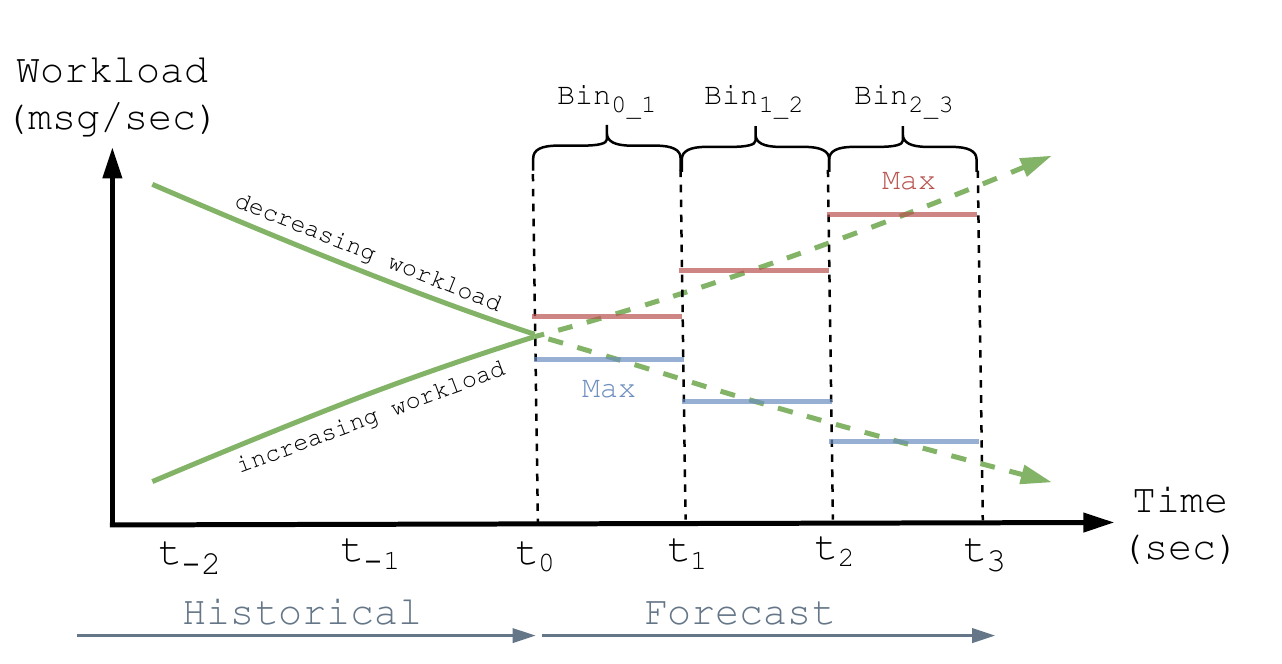}
    \caption{Optimizing for future workloads using averaging bins.}
    \label{fig:bins}
\end{figure}

Once initiated, the online optimization step is performed iteratively based on an \emph{evaluation interval} ($E$).
For each iteration, the job uptime metric is gathered and evaluated.
If the uptime of the targeted DSP job is shorter than $E$, Phoebe makes the assumption that a failure and a rollback recovery has occurred or is currently underway.
In this case Phoebe waits by scheduling the next optimization step to start after $E$ has expired from the current point in time.
This is done to ensure optimizations are not initiated during rollback recoveries and evaluation metrics are not polluted by the recovery process. 
In the case where uptime exceeds $E$, failure-free processing is assumed and the actual optimization procedure can be started.

For any possible scaleout configuration, Phoebe determines its applicability in the current situation with regards to recovery time and latency expectations.
In a first step, Phoebe uses the model $M_R$ and iterates over the full range of possible scaleouts.
By definition, $M_R$ returns $R$ and the projected $T_{avg}$.
The latter is determined as follows: The forecast used within $M_R$ is first separated into a number of averaging bins after which the value for the bin with the maximum average expected workload rate is calculated.
It is this prospective value that is used in place of the $T_{avg}$ for the current optimization step.
This ensures that if the workload is expected to increase, then the furthest bin is selected and optimizations target higher workload rates further away from the current point in time. 
Conversely, if the workload is expected to decrease, the closest bin is selected resulting in optimizations targeting scale-downs closer to the current point in time.
\autoref{fig:bins} shows a graphical overview of this process.
Importantly, this method ensures that resources are never in shortfall and the number of reconfigurations is reduced resulting in high levels of uptime.

After the feasibility of each scaleout is assessed using $R$ and $RC_{tar}$, a second investigation is conducted with respect to latency expectations. 
Using the projected $T_{avg}$, the feasibility of each scaleout can be determined through the model $M_L$.
If the predicted latency belongs to the clustering which is assumed to represent normal latencies, the respective scaleout is valid both in terms of recovery time and latency, and thus considered for reconfiguration.
Eventually, the smallest scaleout satisfying both conditions is selected.
The whole procedure is depicted in~\autoref{alg:whole_procedure}.

\begin{algorithm}
    \small
    \caption{Pseudocode of complete procedure}\label{alg:whole_procedure}
    \begin{algorithmic}
        \Require $I, D, RC_{\max}$
        \Ensure $S_{opt} \in \mathbf{S}$
        \State $S_{opt} \gets \max(\mathbf{S})$
        \ForAll{$S_i \in \mathbf{S}$} \Comment{loop over sorted scaleout options}
        \State $R, T_{avg} \gets M_R(S_i, I, D)$
        \If{$R \leq RC_{tar}$} \Comment{validate constraint}
        \State $L_{avg}, L_C \gets M_L(S_i, T_{avg})$
        \If{$L_C < 1$} \Comment{require normal latencies}
        \State $S_{opt} \gets S_i$
        \State \textbf{break}
        \EndIf
        \EndIf
        \EndFor
    \end{algorithmic}
\end{algorithm}

If this result is distinct from the current scaleout, then Phoebe initiates a reconfiguration which ensures the least amount of resources are allocated to deliver the best possible latencies at the projected workload while giving assurances that recovery times are close to the expected $RC_{tar}$ target.

\section{Evaluation}
\label{sec:evaluation}

Now we demonstrate that Phoebe is both practical and beneficial by performing two experiments and presenting a comparison which includes static configuration setups as well as state of the art approaches.
The prototype, data, and experiment artifacts can be found in the following repository\footnote{\url{https://github.com/dos-group/phoebe}}.

\subsection{Experimental Setup}

Our experimental setup consisted of a co-located 5-node Kubernetes~\cite{VPK+15} and HDFS~\cite{SKR+10} cluster with a single switch connecting all servers.
Each experiment consisted of a Kubernetes namespace containing: an Apache Kafka\cite{KNR11} cluster configured with 24 partitions and a replication factor of 3; an Apache Flink\cite{CKE+15} session cluster; and a Prometheus\footnote{\url{https://prometheus.io}, Accessed: Mar 2022} time series database for the gathering of metrics.
All sources and sinks of the experimental processing pipelines were configured to use exactly-once processing thereby guaranteeing the consistency of results.
Flink taskmanagers were allocated one full CPU core with 2048 MB of memory.
Node specifications and software versions are summarized in~\autoref{tbl:clusterspecs}.
Regarding end-to-end latencies, averages were measured over a 2 minute windowing period using the $95^{th}$ percentile in order to filter outliers during normal failure-free operations. 
Experiments were designed to run for 6 hours. 
Each experiment was conducted 5 times with the median selected for our results and discussion.
In order to evaluate recovery times, Chaos Mesh\footnote{\url{https://chaos-mesh.org}, Accessed: Mar 2022} was used to inject a total of 8 timeout failures per experimental run at an interval of 20 minutes.
Doing so ensured an even distribution of failures across a wide range of workload rates.

\begin{table}
\centering
\caption{Cluster Specifications}
\begin{tabular}[t]{rp{0.65\linewidth}}
    \toprule
    Resource&Details\\
    \midrule
    OS&Ubuntu 20.04.1\\
    CPU&AMD EPYC 7282 16-Core Processor, 32 cores, 2.8 GHz\\
    Memory&128 GB RAM\\
    Storage&2TB RAID0 (2x1TB SSD, linux software RAID)\\
    Network&10 GBit Ethernet NIC\\
    Software&Java v1.11, Flink v1.14, Kafka v2.8, ZooKeeper v3.6, Docker v19.3, Kubernetes v1.24, HDFS v2.8, Redis v5.0, Prometheus v2.25, Chaos Mesh v2.1
    \\
    \bottomrule
\end{tabular}
\label{tbl:clusterspecs}
\end{table}
\setlength{\textfloatsep}{0.1cm}

\subsection{Phoebe Setup}
For profiling, the DSP job was configured to execute for 7 minutes at each selected workload rate allowing 5 minutes to pass for conditions to normalize after each workload change before latency measurements were taken.
Profiling runs were initialized at 20K msg/sec with step increments of 20K.
Phoebe was configured to profile across a total of 8 parallel deployments with sample scaleouts equidistantly spaced from within the range of 2 to 24.
During optimization, the evaluation interval was configured to execute every 10 minutes.
Likewise, a 10 minute time horizon $H_{max}$ was used for the TSF model.
From observations, we determined that the shortest achievable recovery times even under low workloads was in the region of 120s.
Based on this, for both experiments a reasonable QoS recovery time target of 180s was chosen for optimization.

\subsection{Baseline Setup}

\subsubsection{Static Configurations}

For the purposes of our experiments we included 3 baseline executions consisting of static configurations.
Selected scaleouts included 4, 12, and 24 workers.
This setup is intended to showcase how performance is impacted when the resources of the DSP jobs are such that they fall between the ranges of under- to over-provisioning.

\subsubsection{Flink Reactive}

The reactive mode scheduler\footnote{\url{https://flink.apache.org/2021/05/06/reactive-mode}, Accessed: Mar 2022} allows Apache Flink to react to resources being added or removed from the cluster.
This mode ensures that all workers assigned to the cluster will be utilized when changes occur and automatically restarts the DSP job from the last successful checkpoint with the updated scaleout.
When used in combination with a utility to monitor specific metrics and automatically scale the set of resources, Apache Flink is provided with a built-in autoscaling solution.
For the purposes of these experiments, we made use of the Kubernetes Horizontal Pod Autoscaler\footnote{\url{https://kubernetes.io/docs/tasks/run-application/horizontal-pod-autoscale}, Accessed: Mar 2022}.
As recommended, the timeout interval for Flink taskmanagers was set to 20s with the checkpoint interval set to 10s.
To ensure results were comparable, all experiments were configured as such.
The Horizontal Pod Autoscaler was configured to target CPU 35\% utilization, the same as in the exemplary setup. 

\subsubsection{TWRES}

We employ a second dynamic scaling baseline inspired from recent related work. 
Precisely, we use the resource scaling algorithm (\emph{TWRES}) proposed in~\cite{HuKZ19} for Spark streaming jobs.
Similar to Phoebe, this algorithm requires profiling data, and scales a data processing application under consideration of workload forecasts, a performance model for the maximum processing capacity of individual scaleouts, and formulated latency constraints.
At every evaluation interval, it chooses the smallest scaleout which is still able to cope with the prospective future throughput, based on its estimated maximum processing capacity. 
If the current scaleout is expected to be valid and thus a downscaling can be considered, the current latency is compared to the latency constraint, and an increment by one of the current scaleout is conducted in case of violation.
For a fair comparison, TWRES uses the same profiling data, workload forecasting technique, and evaluation interval as Phoebe, such that the actual rescaling approaches can be investigated in detail. 
Since required by TWRES, we define a reasonable latency constraint of 2000ms, which we derived from our observations during the profiling phase.

\subsection{Streaming Jobs}

\subsubsection{Top Speed Windowing (TSW) Experiment}

For the first experiment, a DSP job was selected from the official Flink repository\footnote{\url{https://github.com/apache/flink/}; Accessed: Mar 2022}.
It was modified so that sources consumed events from and sinks published results to separate Apache Kafka topics.
This job showcases grouped stream windowing where different eviction and trigger policies can be used. 
A source fetches car events containing a unique number-plate, their current speed (km/h), overall elapsed distance (m) and a timestamp. 
The streaming job triggers the top speed of each car every 50 meters elapsed for the last 10 seconds.
The number of concurrent vehicles over time, i.e. the workload, was created using Sumo~\cite{SUMO2018} and specifically based on the TAPASCologne scenario\footnote{\url{https://sumo.dlr.de/docs/}; Accessed: Mar 2022}.
We designed an application which generated car events limited by this workload rate.
A representation of the number of concurrent cars over time can be seen in \autoref{fig:evaluation_datasets}(a).

\subsubsection{Yahoo Streaming Benchmark (YSB) Experiment}

\begin{figure*}
\centering
\subfloat[TSW experiment: Throughput rates and reconfigurations.]{
  \includegraphics[width=\columnwidth]{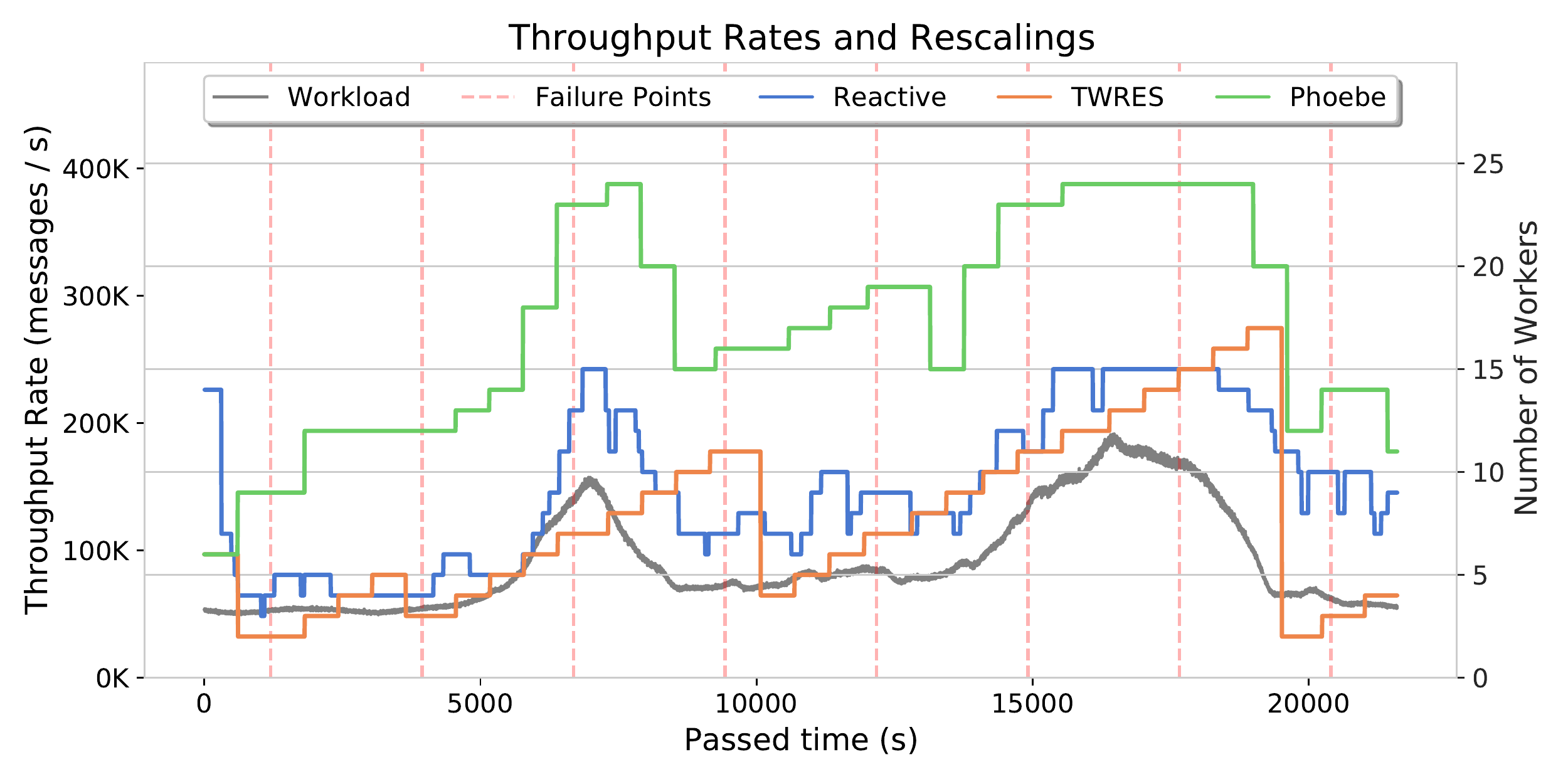}
}
\subfloat[YSB experiment: Throughput rates and reconfigurations.]{
  \includegraphics[width=\columnwidth]{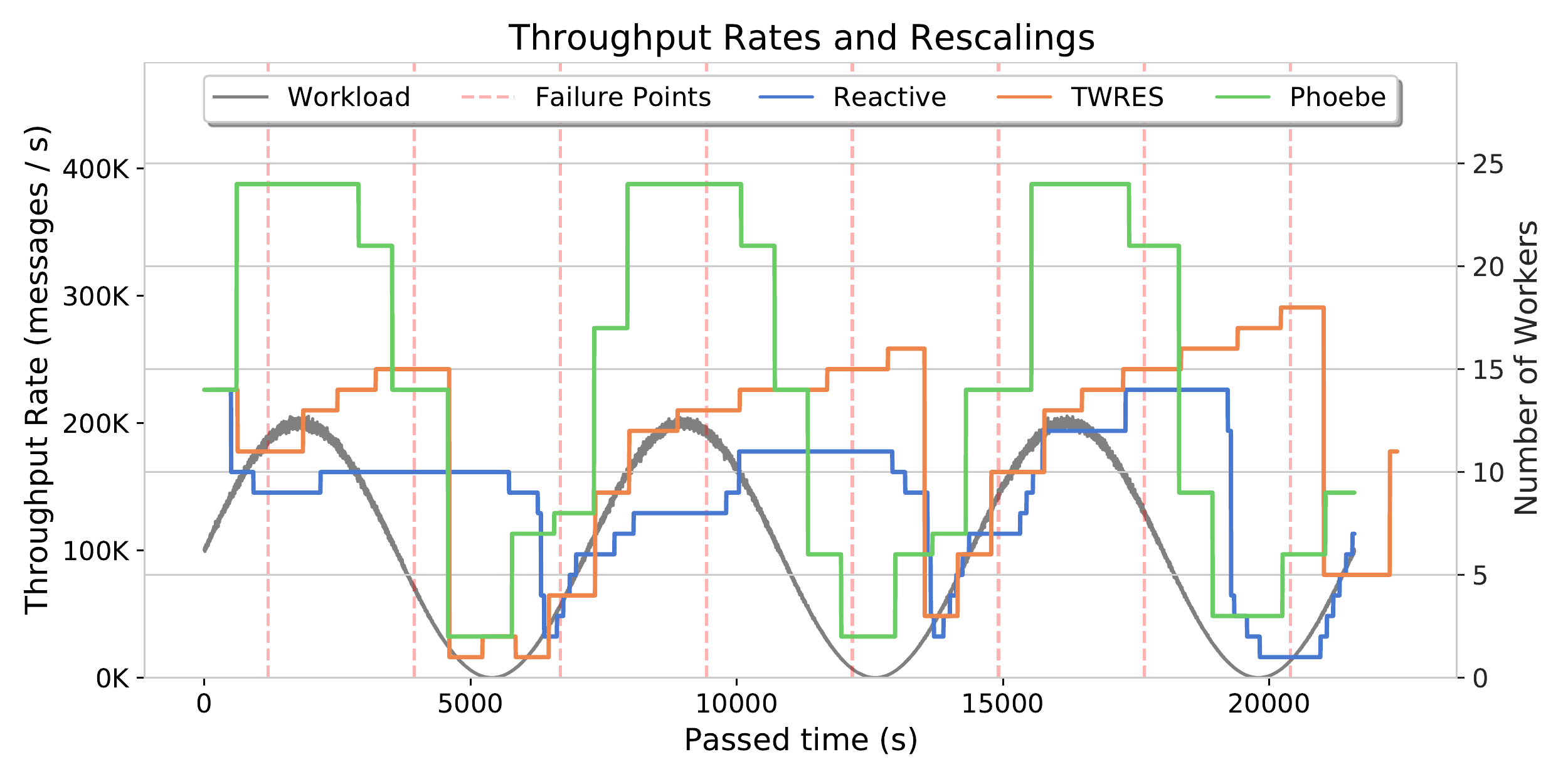}
}
\newline
\subfloat[TSW experiment: Average end-to-end latencies.]{
  \includegraphics[width=\columnwidth]{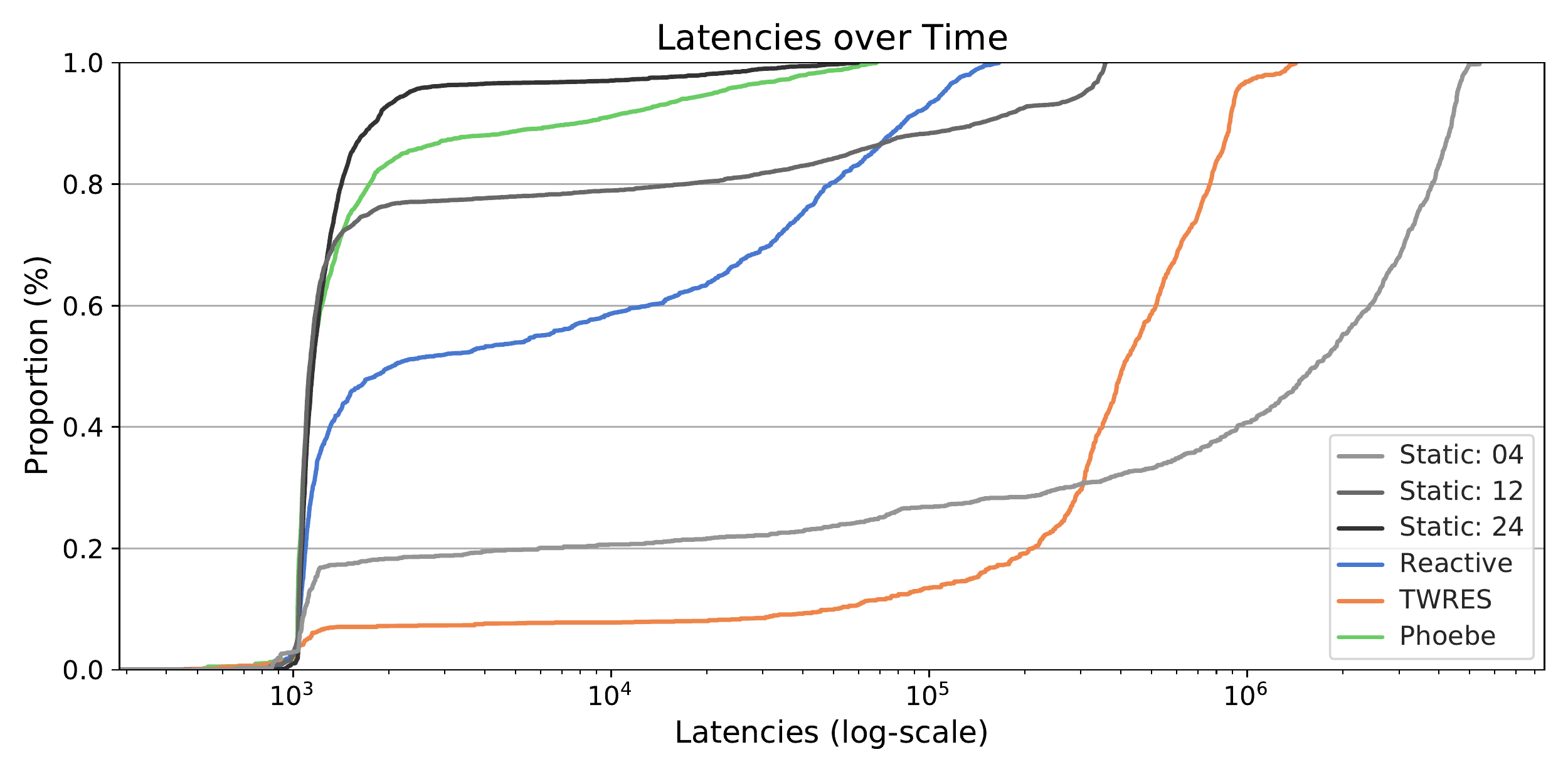}
}
\subfloat[YSB experiment: Average end-to-end latencies.]{
  \includegraphics[width=\columnwidth]{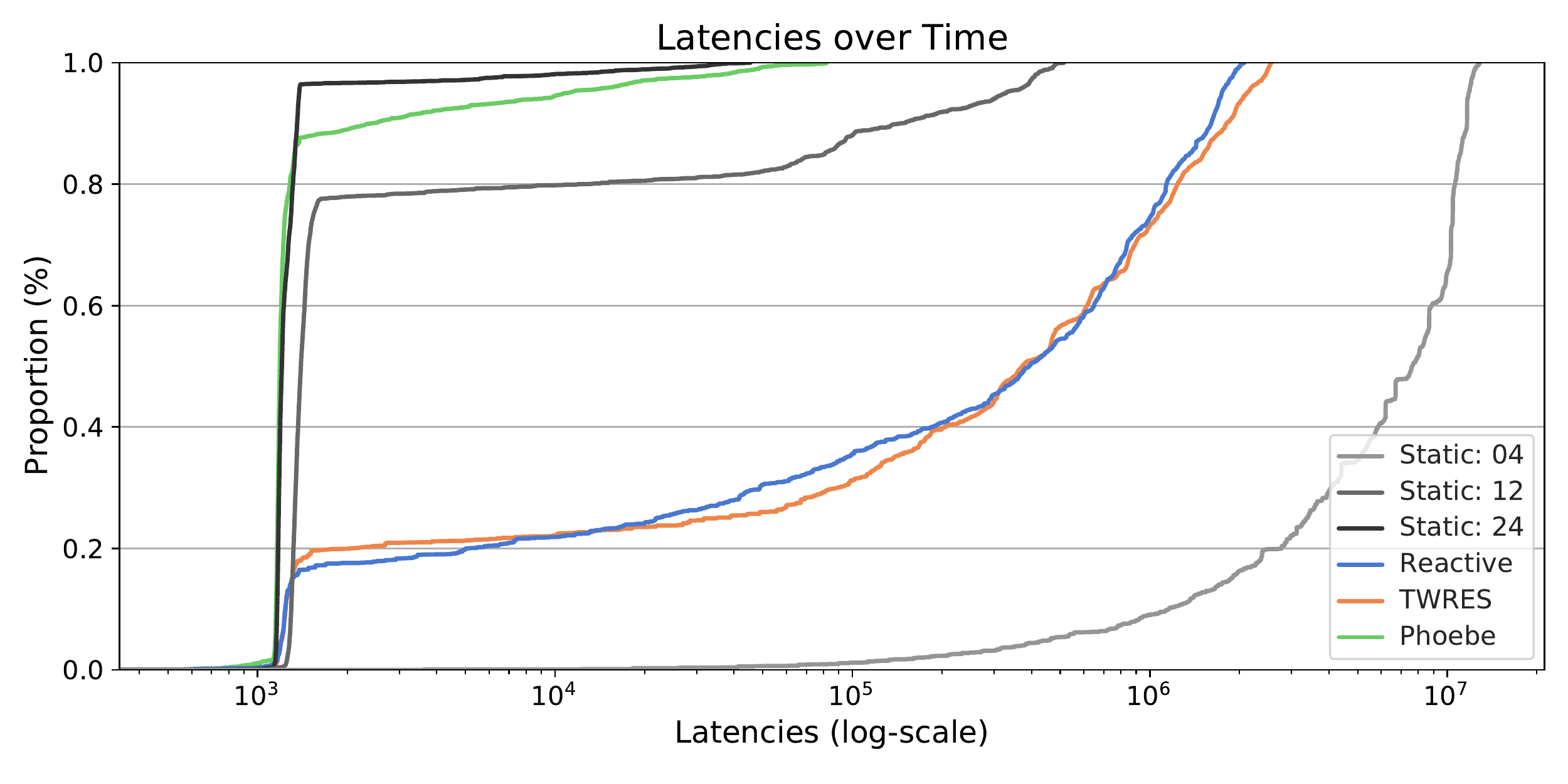}
}
\newline
\subfloat[TSW experiment: Resource consumption over duration of experiment.]{
  \includegraphics[width=\columnwidth]{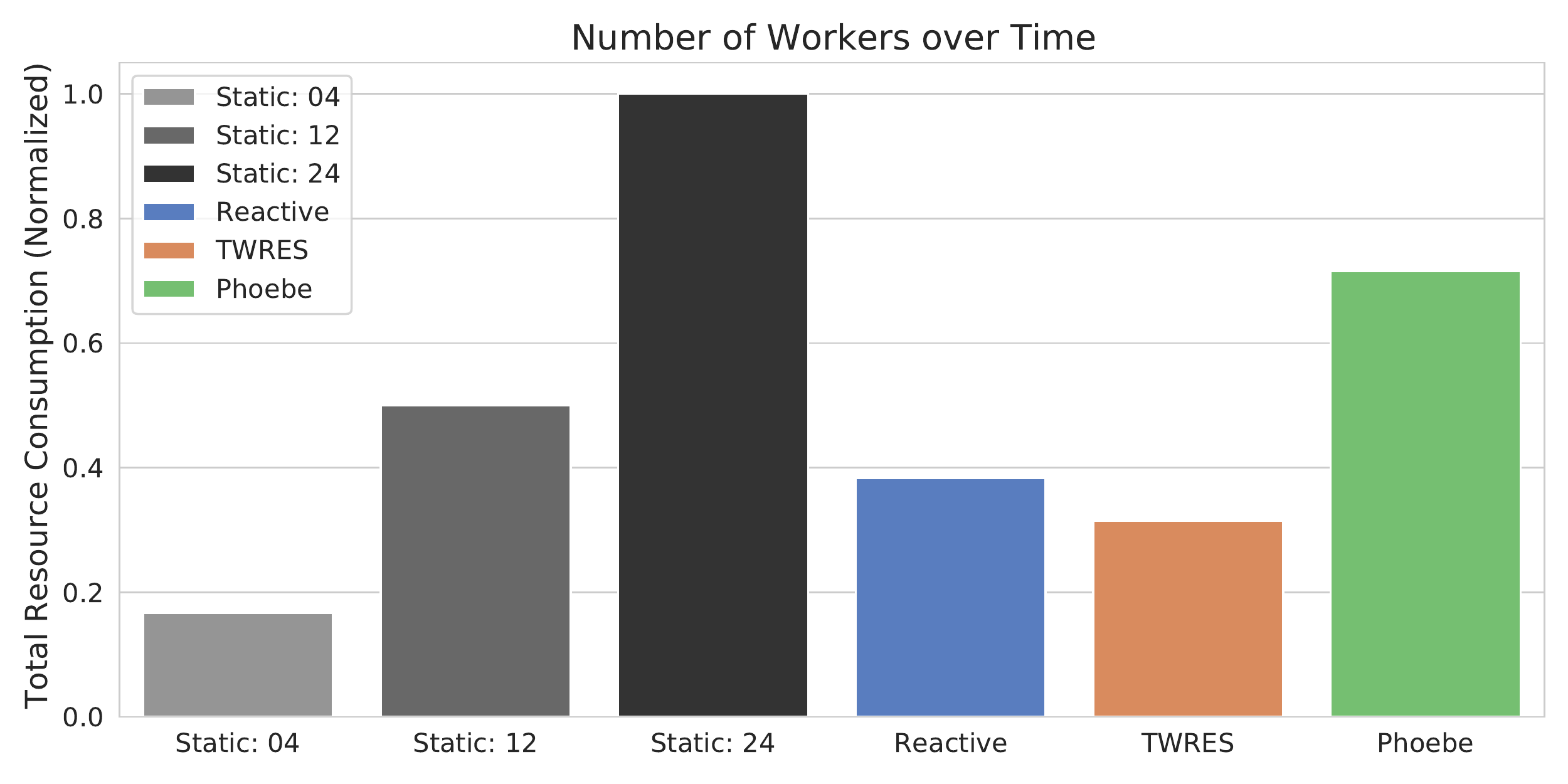}
}
\subfloat[YSB experiment: Resource consumption over duration of experiment.]{
  \includegraphics[width=\columnwidth]{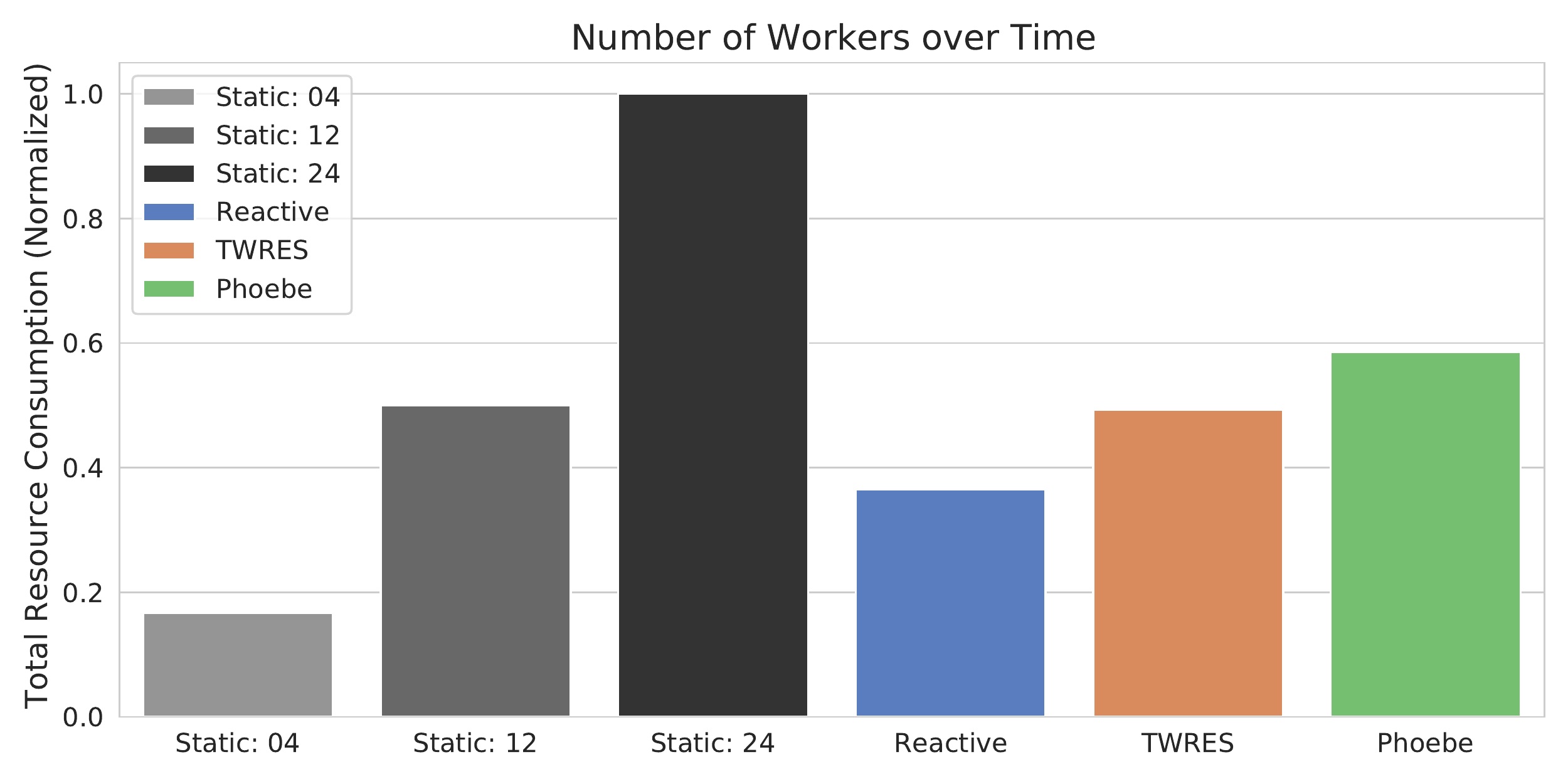}
}
\caption{Overview of performance comparison results for Phoebe and State of the Art Approaches.}
\label{fig:evaluation_datasets}
\end{figure*}

Our second experiment was based on the Yahoo Streaming Benchmark\footnote{\url{https://github.com/yahoo/streaming-benchmarks/}, Accessed: Mar 2022}. 
It implements a streaming advertisement job where there are a number of advertising campaigns and multiple advertisements per campaign.
Streaming sources read from a Kafka topic, identify relevant events, and store a windowed count of these events by campaign. 
In addition to the general setup, this experiment required the deployment of a Redis\footnote{\url{https://redis.io/}, Accessed: Mar 2022} cluster.
For the purpose of this experiment, we modified the benchmark by enabling checkpointing and replacing the handwritten windowing functionality with the default Flink implementation.
A Kafka Producer application was created that would generate a constant stream of events containing, among other things, an $event\_time$, an $event\_type$, and an $ad\_id$. 
The number of msg/sec was limited by a sinusoidal function, similar to what was used in the Flink reactive mode example setup.
A variance of positive and negative 10\% was applied to the number of msg/sec from the generators.
A representation of this workload can be seen in~\autoref{fig:evaluation_datasets}(b).

\subsection{Experimental Results \& Discussion}

At the conclusion of all experiments, metrics were collated and analyzed.
Figures \ref{fig:evaluation_datasets}(a) and \ref{fig:evaluation_datasets}(b) present a graphical representation of the incoming workloads rates over time together with an overlay of the scaleout selections for the 3 autoscaling approaches.
Phoebe is characterized by higher scaleouts which is to be expected as additional resources are allocated for recovery.
In the first part of our analysis we evaluate recovery times in response to the 8 failure injections.
We accomplished this by inspecting the average end-to-end latency metrics and determining the length of times required for values to once again return to normal after rollback recoveries were completed.
Tables \ref{tbl:recovery_times}(a) and \ref{tbl:recovery_times}(b) present an overview of these measurements for each experiment as well as the workload rates at which the failures occurred.
As the static 24-worker configuration was allocated the maximum resources, it produced the fastest recovery times and therefore the benchmark against which all others should be compared.
In instances where recovery times were in excess of two times the optimal target, i.e. 6 minutes, we indicate this with $6m+$.

From these measurements we can conclude that for both experiments, the static 4-worker configuration, reactive, and TWRES experienced resource shortfalls and therefore performed poorly.
The static 12-worker configuration was able to recover in most cases, however, when the workload increased above a certain rate, it too performed poorly.
Phoebe, on the other hand, was the only other method to achieve comparable results and showed the ability to recover reliably at any workload rate.
An error analysis revealed that, for Phoebe, recovery times were on average within 4.5\% of the expected for the TSW experiment, and within 8.5\% of the expected for the YSB experiment.
From this we can conclude that Phoebe was able to provide adequate resources when required for the DSP job to recover in a timeous manner based on the 180s target.

\begin{table}
\caption{Recovery times and reconfigurations.}
\centering
    \subfloat[TSW experiment results.] {
    \begin{tabular}{p{10mm}|p{4mm}p{4mm}p{4mm}p{4mm}p{4mm}p{4mm}p{4mm}p{4mm}|p{4mm}} \toprule
        {} & {$\#1$} & {$\#2$} & {$\#3$} & {$\#4$} & {$\#5$} & {$\#6$} & {$\#7$} & {$\#8$} & {$\Delta$} \\ \midrule
        {Workload} & 52K & 54K & 140K & 73K & 84K & 135K & 166K & 62K & --\\ \midrule[1pt]
        {Static 24} & \hl{140s} & \hl{150s} & \hl{200s} & \hl{165s} & \hl{155s} & \hl{185s} & \hl{240s} & \hl{160s} & --\\ \midrule
        {Static 12} & 160s & 165s & 6m+ & 195s & 205s & 6m+ & 6m+ & 210s & --\\ \midrule
        {Static 4} & 6m+ & 6m+ & 6m+ & 6m+ & 6m+ & 6m+ & 6m+ & 6m+ & --\\ \midrule
        {Phoebe} & 145s & 170s & \hl{200s} & 175s & 175s & 210s & \hl{240s} & 210s & 20 \\ \midrule
        {Reactive} & 6m+ & 6m+ & 6m+ & 6m+ & 6m+ & 6m+ & 6m+ & 6m+ & 63 \\ \midrule
        {TWRES} & 6m+ & 6m+ & 6m+ & 6m+ & 6m+ & 6m+ & 6m+ & 135s & 29 \\ \midrule\bottomrule[1pt]
    \end{tabular}
    }
    \newline
    \subfloat[YSB experiment results.] {
    \begin{tabular}{p{10mm}|p{4mm}p{4mm}p{4mm}p{4mm}p{4mm}p{4mm}p{4mm}p{4mm}|p{4mm}} \toprule
        {} & {$\#1$} & {$\#2$} & {$\#3$} & {$\#4$} & {$\#5$} & {$\#6$} & {$\#7$} & {$\#8$} & {$\Delta$} \\ \midrule
        {Workload} & 186K & 72K & 60K & 195K & 7K & 136K & 130K & 14K & --\\ \midrule[1pt]
        {Static 24} & \hl{195s} & \hl{155s} & \hl{155s} & \hl{190s} & \hl{135s} & \hl{185s} & \hl{180s} & \hl{145s} & --\\ \midrule
        {Static 12} & 6m+ & 210s & 195s & 6m+ & 145s & 6m+ & 290s & 155s & --\\ \midrule
        {Static 4} & 6m+ & 6m+ & 6m+ & 6m+ & 6m+ & 6m+ & 6m+ & 6m+ & --\\ \midrule
        {Phoebe} & 235s & 180s & 195s & 215s & 160s & 330s & \hl{180s} & 160s & 20 \\ \midrule
        {Reactive} & 6m+ & 6m+ & 6m+ & 6m+ & 6m+ & 6m+ & 6m+ & 6m+ & 40 \\ \midrule
        {TWRES} & 6m+ & 6m+ & 6m+ & 350s & 230s & 6m+ & 340s & 6m+ & 24 \\ \midrule\bottomrule[1pt]
    \end{tabular}
    }
    \label{tbl:recovery_times}
\end{table}
\setlength{\textfloatsep}{0.1cm}

Next we evaluate overall performance in terms of average end-to-end latencies.
Figures \ref{fig:evaluation_datasets}(c) and \ref{fig:evaluation_datasets}(d) shows an \emph{empirical cumulative distribution function} for all static and dynamic-scaling baselines.
Unsurprisingly, the static 24-worker setup performed best with 90\% of latency measurements falling within the near-optimal range of approximately 1000ms.
The distribution of latencies falling outside of this are caused by the injected failures and subsequent rollback recoveries.
Phoebe was the second best performer with 80\% of latency measurements showing the same near-optimal performance.
This roughly 10\% difference can be attributed to reconfigurations, however, Phoebe was able to achieve stable performance for both experiments.

The static 12-worker setup was the third best performer.
For the remaining static 4-worker setup and dynamic scaling approaches, performance was significantly worse.
Reactive and TWRES were only able to achieve near-optimal latencies for less than 20\% of all measurements, apart from the TSW experiment where reactive achieved 35\%.
From this we can conclude that they were unable to deliver a stable service.
A major contributing factor for this can be seen in Tables \ref{tbl:recovery_times}(a) and \ref{tbl:recovery_times}(b) where the $\Delta$ symbol represents the number of configuration changes initiated over the course of the experiment.
A larger number of changes contributes to higher latencies as a result of restarts and likewise indicates that scaling decisions were imprecise and short lived.

Finally we evaluate resource utilization.
Figures \ref{fig:evaluation_datasets}(e) and \ref{fig:evaluation_datasets}(f) show the cumulative costs of each experiment over 6 hours.
We calculate cost by summing the total number of containers per second for each approach and normalizing the results for comparison.
We have already concluded that only the static 24-worker setup and Phoebe were able to produce near-optimal performance as well as reliable recovery, therefore we focus exclusively on these.
A direct comparison shows that Phoebe was able to provide significant improvements over the static 24-worker setup with savings of 25.5\% and 41\% in resource utilization.
When profiling costs are added, this saving decreases to 2.5\% and 9.9\% for each experiment.
However, it is important to note that total resource utilization with profiling included is directly related to the duration of the experiment.
Because profiling is only executed once, if the experiments were to continue executing and assuming similar workload behaviors, utilization would improve over time as the static profiling costs become less significant.
Therefore, in essence Phoebe is able to buy back the cost of profiling within the first 6 hours of operation and utilization would improve by 22\% and 33.5\% after the first day, and finally tend towards the reported 25.5\% and 41\% after one week.

\section{Related Work}
\label{sec:sota}

In this section we examine work related to automatic system tuning for stateful distributed dataflow jobs through the elastic scaling of compute nodes.
The majority of approaches rely on course-grained metrics to make scaling decisions usually involving monitoring for bottlenecks and fixed/percentage-based resource adjustments.
Flink and Spark Streaming, i.e. the two  most popular DSP frameworks, both provide autoscaling solutions which fit into this category.
In \cite{GSH+14}, Gedik et al. propose a solution for IBM Infosphere Streams~\cite{Biem2010IBMIS} where backpressure and congestion are observed.
Likewise, Dhalion~\cite{FAG+17} provides a similar service for Heron.
Here scaling decisions are based on the status of individual operators which can results in long convergence times as each reconfiguration targets a single operator.
Also, backpressure can be an unreliable metric as it is susceptible to data skew.
In \cite{PETROV2018109}, Petrov et al. present a performance model where scaling decisions are derived from processing latencies.
TWRES~\cite{HuKZ19} similarly manages resource allocations by monitoring for violations of user-defined latency constraints.
Both approaches, however, are specific to Spark Streaming with \cite{PETROV2018109} also requiring a modified version of the framework.
Like backpressure, latency metrics tend not to be reliable for triggering changes as they are highly influenced by the transient nature of a shared cluster and the unique characteristics of each individual DSP job.
TWRES is the only other approach we are aware of that uses TSF. 
In the original paper, however, they propose a maximum time horizon of 5 seconds, which does not offer much insight when exactly-once processing guarantees are required.

Still further approaches attempt to model the scaleout behaviours of distributed dataflows.
Much of this is related to batch processing where the scaleout of resources is determined by runtime targets.
In a manner similar to profiling, some approaches model scaleout behaviours based on previous executions of jobs in order to make predictions about resource utilization and completion times. 
Ernest~\cite{Venkataraman2016ErnestEP} allocates cloud resources by running the job on a subset of the inputs and different sets of resources. 
Bell~\cite{Thamsen2016SelectingRF} circumvents these isolated training runs and uses available workload data of recurring jobs to make predictions. 
Enel~\cite{ScheinertZTGWAK21}, a context-aware and graph-based modeling approach, follows a more fine-grained strategy as it operates in-between synchronization barriers of a dataflow job and incorporates runtime metrics.
CherryPick~\cite{Alipourfard2017CherryPickAU} uses Bayesian Optimization to build performance modeling for jobs which are just accurate enough to distinguish near-optimal configurations.
This research has expanded into the area of stream processing where the authors of~\cite{KLH+18} present an automatic scaling controller for dynamic workloads which proposes a general performance model to estimate the processing and output rates of individual dataflow operators.
Although they are able to demonstrate that their approach converges quickly and achieves stable performance, it requires a modified version of the DSP framework and does not take exactly-once processing or recovery times into consideration.

Overall, we differentiate our approach from the related work by providing a solution which: is generalizable without requiring a customized version of the DSP framework; automatically selects scaleouts which produces near-optimal end-to-end latencies without requiring input from the user regarding thresholds; uses TSF to anticipate future workload requirements, and; takes exactly-once processing guarantees as well as recovery time planning into consideration.

In our previous work, we also investigated parameter auto-tuning of DSP jobs to improve end-to-end latencies and recovery time, yet focused on optimizing checkpoint intervals while assuming scaleouts to be static~\cite{Geldenhuys2019EffectivelyTS, Geldenhuys2020ChironOF}.
\section{Conclusion}
\label{sec:conclusion}

In this paper we presented Phoebe, an approach which uses TSF to optimize the resource utilization of DSP jobs executing in environments where the workload is expected to change over time.
Not only does it demonstrate the ability to produce near-optimal latencies while reducing resource over-provisioning, but likewise it provides a mechanism for approximating recovery times inline with defined QoS targets.
Phoebe is applicable to scenarios where results are expected to be consistent in the presence of partial failures, i.e. where exactly-once processing is guaranteed.
It achieves this by performing parallel profiling runs, training runtime models on the results, and executing a runtime optimization step where near-optimal scaleouts are selected.
Through experimentation we showed that Phoebe is able to deliver stable end-to-end latencies, something two state of the art methods were unable to achieve, while at the same time providing up to 25.5\% and 41\% better resource utilization in comparison to an over-provisioned setup.
We additionally demonstrate how our recovery times estimation heuristic is able to achieve results which are on average within 4.5\% and 8.5\% of expected.

\section*{Acknowledgment}
This work has been supported through grants by the German Ministry for Education and Research (BMBF) as BIFOLD (funding mark 01IS18025A) and WaterGridSense 4.0 (funding mark 02WIK1475D).

\bibliographystyle{IEEEtran}
\bibliography{paper}
\end{document}